\documentclass{bioinfo}
\copyrightyear{2015}
\pubyear{2015}

\usepackage{amsmath}
\usepackage[ruled,vlined]{algorithm2e}

\SetCommentSty{mycommfont}
\SetKwComment{Comment}{$\triangleright$\ }{}

\usepackage{natbib}
\begin{document}
\firstpage{1}

\title[Long-read mapping and assembly]{Minimap and miniasm: fast mapping and de novo assembly for noisy long sequences}
\author[Li]{Heng Li}
\address{Broad Institute, 75 Ames Street, Cambridge, MA 02142, USA}
\maketitle

\begin{abstract}

\section{Motivation:} Single Molecule Real-Time (SMRT) sequencing technology and Oxford
Nanopore technologies (ONT) produce reads over 10kbp in length, which have
enabled high-quality genome assembly at an affordable cost.  However, at
present, long reads have an error rate as high as 10--15\%.  Complex and
computationally intensive pipelines are required to assemble such reads.

\section{Results:} We present a new mapper, minimap, and a \emph{de novo}
assembler, miniasm, for efficiently mapping and assembling SMRT and ONT reads
without an error correction stage. They can often assemble a sequencing run of
bacterial data into a single contig in a few minutes, and assemble 45-fold
\emph{C. elegans} data in 9 minutes, orders of magnitude faster than the
existing pipelines, though the consensus sequence error rate is as high as raw
reads. We also introduce a pairwise read mapping format (PAF) and
a graphical fragment assembly format (GFA), and demonstrate the
interoperability between ours and current tools.

\section{Availability and implementation:} https://github.com/lh3/minimap and
https://github.com/lh3/miniasm

\section{Contact:} hengli@broadinstitute.org

\end{abstract}

\section{Introduction}

High-throughput short-read sequencing technologies, such as Illumina, have
empowered a variety of biological researches and clinical applications that
would not be practical with the older Sanger sequencing. However, the short
read length (typically a few hundred basepairs) has posed a great challenge to
\emph{de novo} assembly as many repetitive sequences and segmental duplications
are longer than the read length and can hardly be resolved by short reads even
with paired-end data~\citep{Alkan:2011zr}. Although with increased read length
and improved algorithms we are now able to produce much better short-read
assemblies than a few years ago, the contiguity and completeness of the
assemblies are still not as good as Sanger assemblies~\citep{Chaisson:2015wj}.

The PacBio's SMRT technology were developed partly as an answer to the
problem with short-read \emph{de novo} assembly. However, due to the high
per-base error rate, around 15\%, these reads were only used as a complement to
short reads initially~\citep{Bashir:2012gb,Ribeiro:2012bx,Koren:2012pt},
until~\citet{Chin:2013qr} and~\citet{Koren:2013fc} demonstrated the feasibility
of SMRT-only assembly. Since then, SMRT is becoming the preferred technology
for finishing small genomes and producing high-quality Eukaryotic
genomes~\citep{Berlin:2015xy}.

Oxford Nanopore Technologies (ONT) has recently offered another long-read
sequencing technology. Although the per-base error rate was high at the
early access phase~\citep{Quick:2014uf}, the latest data quality has been
greatly improved. \citet{Loman:2015xu} confirmed that we can achieve
high-quality bacterial assembly with ONT data alone.

Published long-read assembly pipelines all include four stages: (i) all-vs-all
raw read mapping, (ii) raw read error correction, (iii) assembly of error
corrected reads and (iv) contig consensus polish. Stage (iii) may involve
all-vs-all read mapping again, but as the error rate is much reduced at this
step, it is easier and faster than stage (i). Table~\ref{tab:tools} shows the tools used for
each stage. Notably, our tool minimap is a raw read overlapper and miniasm is
an assembler. We do not correct sequencing errors, but instead directly produce
unpolished and uncorrected contig sequences from raw read overlaps. The idea of
correction-free assembly was inspired by talks given by Gene Myers.
Sikic et al (personal communication) are also independently exploring such an
approach.

\begin{table}[b]
\processtable{Tools for noisy long-read assembly}
{\footnotesize\label{tab:tools}
\begin{tabular}{p{2.4cm}p{2cm}l}
\toprule
Functionality & Program & Reference \\
\midrule
Raw read overlap & BLASR & \citet{Chaisson:2012aa}\\
& DALIGNER & \citet{DBLP:conf/wabi/Myers14} \\
& MHAP & \citet{Berlin:2015xy} \\
& GraphMap & \citet{sovic:2015aa} \\
& minimap & this article \\
Error correction & pbdagcon & http://bit.ly/pbdagcon \\
& falcon\_sense & http://bit.ly/pbfcasm \\
& nanocorrect & \citet{Loman:2015xu} \\
Assembly & wgs-assembler & \citet{Myers:2000kl} \\
& Falcon & http://bit.ly/pbfcasm \\
& ra-integrate & http://bit.ly/raitgasm \\
& miniasm & this article \\
Consensus polish & Quiver & http://bit.ly/pbquiver \\
& nanopolish & \citet{Loman:2015xu} \\
\botrule
\end{tabular}
}{}
\end{table}

As we can see from Table~\ref{tab:tools}, each stage can be achieved with multiple tools.
Although we have successfully combined tools into different pipelines, we
need to change or convert the input/output formats to make them work
together. Another contribution of this article is the proposal of concise
mapping and assembly formats, which will hopefully encourage modular design of
assemblers and the associated tools.

\begin{methods}
\section{Methods}

\subsection{General notations}

Let $\Sigma=\{\mathrm{A},\mathrm{C},\mathrm{G},\mathrm{T}\}$ be the
alphabet of nucleotides. For a symbol $a\in\Sigma$, $\overline{a}$ is the
Watson-Crick complement of $a$. A string $s=a_1a_2\cdots a_n$ over
$\Sigma$ is also called a \emph{DNA sequence}. Its length is $|s|=n$;
its \emph{reverse complement} is $\overline{s}=\overline{a_1a_2\cdots
a_n}=\overline{a}_n\overline{a}_{n-1}\cdots\overline{a}_1$.
For convenience, we define strand function
$\pi:\Sigma^*\times\{0,1\}\to\Sigma^*$ such that $\pi(s,0)=s$ and
$\pi(s,1)=\overline{s}$. Here $\Sigma^*$ is the set of all DNA sequences.

By convention, we call a $k$-long DNA sequence as a \emph{$k$-mer}. We use the
notation $s^k_i=a_i\cdots a_{i+k-1}$ to denote a $k$-long substring of $s$
starting at $i$. $\Sigma^k$ is the set of all $k$-mers.

\subsection{Minimap}

\subsubsection{Overview of $k$-mer based sequence similarity search}\label{sec:minimapov}

BLAST~\citep{Altschul:1997vn} and BLAT~\citep{Kent:2002jk} are among the most
popular sequence similarity search tools. They use one $k$-mer hash function
$\phi:\Sigma^k\to\mathbb{Z}$ to hash $k$-mers at the positions
$1,w+1,2w+1,\ldots$ of a target sequence and keep the hash values in a hash
table. Upon query, they use the same hash function on every $k$-mer of the
query sequence and look up the hash table for potential matches. If there are
one or multiple $k$-mer matches in a small window, these aligners extend the
matches with dynamic programming to construct the final alignment.

DALIGNER~\citep{DBLP:conf/wabi/Myers14} does not use a hash table. It instead
identifies $k$-mer matches between two sets of reads by sorting $k$-mers and
merging the sorted lists. DALIGNER is fast primarily because sorting and
merging are highly cache efficient.

MHAP~\citep{Berlin:2015xy} differs from others in the use of MinHash
sketch~\citep{Broder:1997aa}.  Briefly, given a read sequence $s$ and $m$
$k$-mer hash functions $\{\phi_j\}_{1\le j\le m}$, MHAP computes
$h_j=\min\{\phi_j(s_i^k):1\le i\le |s|-k+1\}$ with each hash function $\phi_j$, and
takes list $(h_j)_{1\le j\le m}$, which is called the \emph{sketch} of
$s$, as a reduced representation of $s$. Suppose $(h_j)_j$ and $(h'_j)_j$ are
the sketches of two reads, respectively. When the two reads are similar to each
other or have significant overlaps, there are likely to exist multiple $j$ such
that $h_j=h'_j$. Potential matches can thus be identified. A limitation of
MinHash sketch is that it always selects a fixed number of hash values
regardless of the length of the sequences. This may waste space or hurt
sensitivity when input sequences vary greatly in lengths.

Minimap is heavily influenced by all these works. It adopts the idea of sketch
like MHAP but takes minimizers \citep{DBLP:conf/sigmod/SchleimerWA03,Roberts:2004fv} as a reduced
representation instead; it stores $k$-mers in a hash table like BLAT and MHAP
but also uses sorting extensively like DALIGNER. In addition, minimap is
designed not only as a read overlapper but also as a read-to-genome and
genome-to-genome mapper. It has more potential applications.

\subsubsection{Computing minimizers}

\begin{algorithm}[tb]
\DontPrintSemicolon
\footnotesize
\KwIn{Parameter $w$ and $k$ and sequence $s$ with $|s|\ge w+k-1$}
\KwOut{($w$,$k$)-minimizers, their positions and strands}
\BlankLine
\textbf{Function} {\sc MinimizerSketch}$(s,w,k)$
\Begin {
	$\mathcal{M}\gets\emptyset$\Comment*[r]{NB: $\mathcal{M}$ is a set; no duplicates}
	\For{$i\gets1$ \KwTo $|s|-w-k+1$} {
		$m\gets\infty$\;
		\nl\For (\Comment*[f]{Find the min value}) {$j\gets0$ \KwTo $w-1$} {
			$(u,v)\gets(\phi(s^k_{i+j}),\phi(\overline{s}^k_{i+j}))$\;
			\If (\Comment*[f]{Skip if strand ambiguous}) {$u\not=v$} { 
				$m\gets\min(m,\min(u,v))$\;
			}
		}
		\nl\For (\Comment*[f]{Collect minimizers}) {$j\gets0$ \KwTo $w-1$} {
			$(u,v)\gets(\phi(s^k_{i+j}),\phi(\overline{s}^k_{i+j}))$\;
			\uIf{$u<v$ {\bf and} $u=m$} {
				$\mathcal{M}\gets\mathcal{M}\cup\{(m,i+j,0)\}$\;
			}\ElseIf{$v<u$ {\bf and} $v=m$}{
				$\mathcal{M}\gets\mathcal{M}\cup\{(m,i+j,1)\}$\;
			}
		}
	}
	\Return $M$\;
}
\caption{Compute minimizers}\label{alg:minimizer}
\end{algorithm}

Loosely speaking, a $(w,k)$-minimizer of a string is the
smallest $k$-mer in a surrounding window of $w$ consecutive $k$-mers. Formally,
let $\phi:\Sigma^k\to\mathbb{Z}$ be a $k$-mer hash function.
A \emph{double-strand $(w,k,\phi)$-minimizer}, or simply a \emph{minimizer}, of a
string $s$, $|s|\ge w+k-1$, is a triple $(h,i,r)$ such that there exists
$\max(1,i-w+1)\le j\le\min(i,|s|-w-k+1)$ which renders
$$
h=\phi(\pi(s_i^k,r))=\min\big\{\phi(\pi(s_{j+p}^k,r')):0\le p<w,r'\in\{0,1\}\big\}
$$
Let $\mathcal{M}(s)$ be the set of minimizers of $s$.  Algorithm~\ref{alg:minimizer} gives the
pseudocode to compute $\mathcal{M}(s)$ in $O(w\cdot|s|)$ time.  Our actual
implementation is close to $O(|s|)$ in average case. It uses a queue to cache
the previous minimals and avoids the loops at line~1 and~2 most of time. In
practice, time spent on collecting minimizers is insignificant.

A natural choice of hash function $\phi$ is to let $\phi(\mathrm{A})=0$,
$\phi(\mathrm{C})=1$, $\phi(\mathrm{G})=2$ and $\phi(\mathrm{T})=3$ and for a
$k$-mer $s=a_1\cdots a_k$, define
$$
\phi(s)=\phi(a_1)\times4^{k-1}+\phi(a_2)\times4^{k-2}+\cdots+\phi(a_k)
$$
This hash function always maps a $k$-mer to a distinct $2k$-bit integer. A
problem with this $\phi$ is that poly-A, which is often highly enriched in
genomes, always gets zero, the smallest value. We may oversample these
non-informative poly-A and hurt practical performance. To alleviate this issue,
we use function $\phi'=h\circ\phi$ instead, where $h$ is an invertible integer
hash function on $[0,4^k)$ (Algorithm~\ref{alg:invhash}; http://bit.ly/invihgi). The
invertibility of $h$ is not essential, but as such $\phi'$ never maps two
distinct $k$-mers to the same $2k$-bit integer, it helps to reduce hash
collisions.

\begin{algorithm}[tb]
\DontPrintSemicolon
\footnotesize
\KwIn{$p$-bit integer $x$}
\KwOut{hashed $p$-bit integer}
\BlankLine
\textbf{Function} {\sc InvertibleHash}$(x,p)$
\Begin {
	$m\gets2^p-1$\;
	$x\gets(\mbox{\tt\char126}x+(x\mbox{\tt\char60\char60}21))\mbox{ \tt\char38}\mbox{ }m$\;
	$x\gets x\mbox{ \tt\char94}\mbox{ }x\mbox{\tt\char62\char62}24$\;
	$x\gets(x+(x\mbox{\tt\char60\char60}3)+(x\mbox{\tt\char60\char60}8))\mbox{ \tt\char38}\mbox{ }m$\;
	$x\gets x\mbox{ \tt\char94}\mbox{ }x\mbox{\tt\char62\char62}14$\;
	$x\gets(x+(x\mbox{\tt\char60\char60}2)+(x\mbox{\tt\char60\char60}4))\mbox{ \tt\char38}\mbox{ }m$\;
	$x\gets x\mbox{ \tt\char94}\mbox{ }x\mbox{\tt\char62\char62}28$\;
	$x\gets(x+(x\mbox{\tt\char60\char60}31))\mbox{ \tt\char38}\mbox{ }m$\;
	\Return $x$\;
}
\caption{Invertible integer hash function}\label{alg:invhash}
\end{algorithm}

Note that in a window of $w$ consecutive $k$-mers, there may be more than one
minimizers. Algorithm~\ref{alg:minimizer} keeps them all with the loop at line~2. This way, a
minimizer of $s$ always corresponds to a minimizer of $\overline{s}$.

For read overlapping, we use $k=15$ and $w=5$ to find minimizers.

\subsubsection{Indexing}

\begin{algorithm}[tb]
\DontPrintSemicolon
\footnotesize
\KwIn{Set of target sequences $\mathcal{T}=\{s_1,\ldots,s_T\}$}
\KwOut{Minimizer hash table $\mathcal{H}$}
\BlankLine
\textbf{Function} {\sc Index}$(\mathcal{T},w,k)$
\Begin {
	$\mathcal{H}\gets$ empty hash table\;
	\For{$t\gets1$ \KwTo $T$} {
		$\mathcal{M}\gets${\sc MinimizerSketch}$(s_t,w,k)$\;
		\ForEach{$(h,i,r)\in \mathcal{M}$} {
			$\mathcal{H}[h]\gets\mathcal{H}[h]\cup\{(t,i,r)\}$\;
		}
	}
	\Return $\mathcal{H}$\;
}
\caption{Index target sequences}\label{alg:idx}
\end{algorithm}

Algorithm~\ref{alg:idx} describes indexing target sequences. It keeps minimizers of all target
sequences in a hash table where the key is the minimizer hash and the value is
a set of target sequence index, the position of the minimizer and the strand
(packed into one 64-bit integer).

In implementation, we do not directly insert minimizers to the hash table.
Instead, we append minimizers to an array of two 64-bit integers (one for minimizer sequence and one for position) and sort the array after collecting
all minimizers. The hash table keeps the intervals on the sorted array. This
procedure dramatically reduces heap allocations and cache misses, and is
supposedly faster than direct hash table insertion.

\subsubsection{Mapping}

Given two sequences $s$ and $s'$, we say we find a \emph{minimizer hit}
$(h,x,i,i')$ if there exist $(h,i,r)\in\mathcal{M}(s)$ and
$(h,i',r')\in\mathcal{M}(s')$ with $x=r\oplus r'$ ($\oplus$ is the XOR
operator). Here $h$ is the minimizer hash value, $x$ indicates the relative
strand and $i$ and $i'$ are the positions on the two sequences, respectively.
We say two minimizer hits $(h_1,x,i_1,i'_1)$ and $(h_2,x,i_2,i'_2)$ are
\emph{$\epsilon$-away} if 1) $x=0$ and $|(i_1-i'_1)-(i_2-i'_2)|<\epsilon$
or 2) $x=1$ and $|(i_1+i'_1)-(i_2+i'_2)|<\epsilon$. Intuitively,
$\epsilon$-away hits are approximately colinear within a band of width
$\epsilon$ (500bp by default).  Given a set of minimizer hits $\{(h,x,i,i')\}$, we can cluster
$i-i'$ for $x=0$ or $i+i'$ for $x=1$ to identify long colinear matches.
This procedure is inspired by Hough Transformation mentioned
by~\citet{sovic:2015aa}. 

\begin{algorithm}[tb]
\DontPrintSemicolon
\footnotesize
\KwIn{Hash table $\mathcal{H}$ and query sequence $q$}
\KwOut{Print matching query and target intervals}
\BlankLine
\textbf{Function} {\sc Map}$(\mathcal{H},q,w,k,\epsilon)$
\Begin {
	$\mathcal{A}\gets$ empty array\;
	$\mathcal{M}\gets${\sc MinimizerSketch}$(q,w,k)$\;
	\nl\ForEach (\Comment*[f]{Collect minimizer hits}) {$(h,i,r)\in \mathcal{M}$} {
		\ForEach{$(t,i',r')\in \mathcal{H}[h]$} {
			\uIf (\Comment*[f]{Minimizers on the same strand}) {$r=r'$} {
				Append $(t,0,i-i',i')$ to $\mathcal{A}$\;
			} \Else (\Comment*[f]{On different strands}) {
				Append $(t,1,i+i',i')$ to $\mathcal{A}$\;
			}
		}
	}
	Sort $\mathcal{A}=[(t,r,c,i')]$ in the order of the four values in tuples\;
	$b\gets1$\;
	\nl\For (\Comment*[f]{Cluster minimizer hits}) {$e=1$ \KwTo $|\mathcal{A}|$} {
		\If{$e=|\mathcal{A}|$ {\bf or} $\mathcal{A}[e+1].t\not=\mathcal{A}[e].t$ {\bf or} $\mathcal{A}[e+1].r\not=\mathcal{A}[e].r$ {\bf or} $\mathcal{A}[e+1].c-\mathcal{A}[e].c\ge\epsilon$} {
			\nl$\mathcal{C}\gets$ the maximal colinear subset of $\mathcal{A}[b..e]$\;
			Print the left- and right-most query/target positions in $\mathcal{C}$\;
			$b\gets e+1$\;
		}
	}
}
\caption{Map a query sequence}\label{alg:map}
\end{algorithm}

Algorithm~\ref{alg:map} gives the details of the mapping algorithm. The loop at line~1
collects minimizer hits between the query and all the target sequences. The
loop at line~2 performs a single-linkage clustering to group approximately
colinear hits. Some hits in a cluster may not be colinear because two minimizer
hits within distance $\epsilon$ are always $\epsilon$-away. To fix this issue,
we find the maximal colinear subset of hits by solving a longest increasing
sequencing problem (line~3). This subset is the final mapping result. In
practical implementation, we set thresholds on the size of the subset (4 by
default) and the number of matching bases in the subset to filter poor mappings
(100 for read overlapping).

\subsection{Assembly graph}

Two strings $v$ and $w$ may be mapped to each other based on their sequence
similarity. If $v$ can be mapped to a substring of $w$, we say $w$
\emph{contains} $v$. If a suffix of $v$ and a prefix of $w$ can be mapped to
each other, we say $v$ \emph{overlaps} $w$, written as $v\to w$.
If we regard strings $v$ and $w$ as vertices, the overlap relationship defines
a directed edge between them. The \emph{length} of $v\to w$ equals the length
of $v$'s prefix that is not in the prefix-suffix match.

Let $G=(V,E,\ell)$ be a graph without multi-edges, where $V$ is a
set of DNA sequences (vertices), $E$ a set of overlaps between them (edges) and
$\ell:E\to\Re_+$ is the edge length function. $G$ is said to be
\emph{Watson-Crick complete} if i) $\forall v\in V$, $\overline{v}\in V$ and
ii) $\forall v\to w\in E$, $\overline{w}\to\overline{v}\in E$. $G$ is said to
be \emph{containment-free} if any sequence $v$ is not contained in other
sequences in $V$. If $G$ is both Watson-Crick complete and containment-free, it
is an \emph{assembly graph}. By definition, any vertex $v$ has a
\emph{complement vertex} $\overline{v}$ in the graph and any edge $v\to w$ has
a \emph{complement edge} $\overline{w}\to\overline{v}$.  Let
$\mathrm{deg}^+(v)$ be the outdegree of $v$ and $\mathrm{deg}^-(v)$ be the
indegree. It follows that $\mathrm{deg}^-(v)=\mathrm{deg}^+(\overline{v})$.

An assembly graph has the same topology as a string graph~\citep{Myers:2005bh},
though the interpretation of the vertex set $V$ is different. In a string
graph, $V$ is the set of the two ends of sequences, not the set of forward and
reverse-complemented sequences. De Bruijn graph can be regarded as a special
case of overlap graph. It is also an assembly graph.

In an assembly graph, an edge $v\to w$ is \emph{transitive} if there exist
$v\to u$ and $u\to w$. Removing a transitive edge does not affect the
connectivity of the graph. A vertex $v$ is a \emph{tip} if ${\rm deg}^+(v)=0$
and ${\rm deg}^-(v)>0$. The majority of tips are caused by artifacts or missing
overlaps. A \emph{bubble} is a directed acyclic subgraph with a single source
$v$ and a single sink $w$ having at least two paths between $v$ and $w$, and without connecting the rest of the graph. The
bubble is tight if ${\rm deg}^+(v)>1$ and ${\rm deg}^-(w)>1$. A bubble may be
caused by missing overlaps or by variants between haplotypes in multi-ploidy samples or paralogs.
It is preferred to collapse bubbles for high contiguity, though this introduces loss of information.

\subsection{Miniasm}

\subsubsection{Trimming reads}

Raw read sequences may contain artifacts such as untrimmed adapters and
chimaera. The first step of assembly to reduce such artifacts by examining
read-to-read mappings. For each read, miniasm computes per-base coverage based
on good mappings against other reads (longer than 2000bp with at least
100bp non-redundant bases on matching minimizers). It then identifies the
longest region having coverage three or more, and trims bases outside this
region.

\subsubsection{Generating assembly graph}

\begin{figure}[tb]
\centering
\includegraphics[width=.45\textwidth]{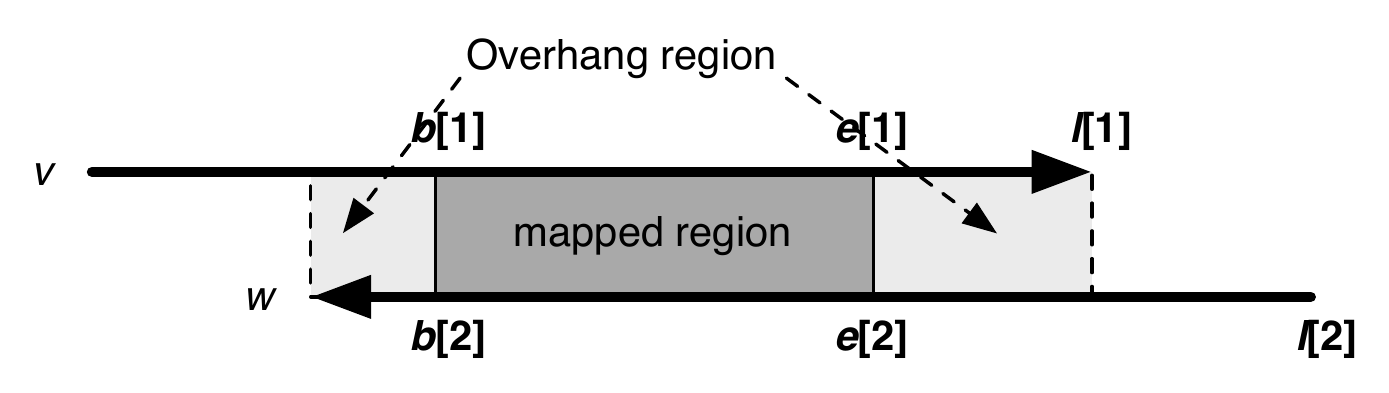}
\caption{Mapping between two reads. $b[1]$ and $e[1]$ are the 0-based starting and
ending mapping coordinates of the first read $v$, respectively. $b[2]$ and
$e[2]$ are the mapping coordinates of read $w$. Lightgray areas indicate
overhang regions that should be mapped together if the overlap is real. If the
overhang regions are small enough, the figure implies an edge $v\to w$ with
approximate length $\ell(v\to w)=b[1]-b[2]$ and its complement edge
$\overline{w}\to\overline{v}$ with
$\ell(\overline{w}\to\overline{v})=(l[2]-e[2])-(l[1]-e[1])$.}\label{fig:overhang}
\end{figure}

\begin{algorithm}[bt]
\DontPrintSemicolon
\footnotesize
\KwIn{Read length $l$, mapping begin coordinate $b$ and mapping end $e$ of the
two reads; max overhang length $o$ (1000 by default) and max overhang to mapping length ratio
$r$ (0.8 by default).}
\KwOut{hashed $p$-bit integer}
\BlankLine
\textbf{Function} {\sc ClassifyMapping}$(l[2], b[2], e[2], o, r)$
\Begin {
	${\it overhang}\gets\min(b[1], b[2])+\min(l[1]-e[1],l[2]-e[2])$\;
	${\it maplen}\gets\max(e[1]-b[1],e[2]-b[2])$\;
	\uIf{${\it overhang}>\min(o,{\it maplen}\cdot r)$} {
		\Return {\tt INTERNAL\_MATCH}
	} \uElseIf {$b[1]\le b[2]$ {\bf and} $l[1]-e[1]\le l[2]-e[2]$} {
		\Return {\tt FIRST\_CONTAINED}
	}\uElseIf {$b[1]\ge b[2]$ {\bf and} $l[1]-e[1]\ge l[2]-e[2]$} {
		\Return {\tt SECOND\_CONTAINED}
	} \uElseIf {$b[1]>b[2]$} {
		\Return {\tt FIRST\_TO\_SECOND\_OVERLAP}
	} \Else {
		\Return {\tt SECOND\_TO\_FIRST\_OVERLAP}
	}
}
\caption{Mapping classification}\label{alg:ovlp}
\end{algorithm}

For each trimmed mapping, miniasm applies Algorithm~\ref{alg:ovlp} to classify the mapping
(see also Figure~\ref{fig:overhang} for the explanation of input variables).
It ignores internal matches, drops contained reads and adds overlaps to the
assembly graph. For a pair of reads, miniasm uses the longest overlap only to
avoid multi-edges.

\subsubsection{Graph cleaning}

\begin{algorithm}[tb]
\DontPrintSemicolon
\footnotesize
\KwIn{$G=(V,E)$, starting vertex $v_0$ and maximum probe distance $d$}
\KwOut{the sink vertex of a bubble within $d$; or {\bf nil} if not found}
\BlankLine
\textbf{Function} {\sc DetectBubble}$(V,E,v_0,d)$
\Begin {
	\lIf{$\mathrm{deg}^+(v_0)<2$} { \Return {\bf nil} } \Comment*[r]{Not a source of bubble}
	\lFor{$v\in V$} { $\delta[v]\gets\infty$ } \Comment*[r]{the min distance from $v_0$ to $v$}
	$\delta[v_0]\gets0$\;
	$S\gets$ empty stack \Comment*[r]{Vertices with all incoming edges visited}
	{\sc Push}$(S,v_0)$\;
	$p\gets0$ \Comment*[r]{Number of visited vertices never added to $S$}
	\While{$S$ is not empty} {
		$v\gets$ {\sc Pop}$(S)$\;
		\ForEach{$v\to w\in E$} {
			\If (\Comment*[f]{A circle involving the starting vertex}) {$w=v_0$} {
				\Return {\bf nil}\;
			}
			\If (\Comment*[f]{Moving too far}) {$\delta[v]+\ell(v\to w)>d$} {
				\Return {\bf nil}\;
			}
			\If (\Comment*[f]{Not visited before}) {$\delta[w]=\infty$} {
				$\gamma[w]\gets \mathrm{deg}^-(w)$ \Comment*[r]{No. unvisited incoming edges}
				$p\gets p+1$\;
			}
			\If{$\delta[v]+\ell(v\to w)<\delta[w]$} {
				\nl$\delta[w]\gets \delta[v]+\ell(v\to w)$\;
			}
			$\gamma[w]\gets\gamma[w]-1$\;
			\If (\Comment*[f]{All incoming edges visited}) {$\gamma[w]=0$} {
				\If (\Comment*[f]{Not a tip}) {$\mathrm{deg}^+(w)\not=0$} {
					{\sc Push}$(S,w)$\;
				}
				$p\gets p-1$\;
			}
		}
		\If (\Comment*[f]{Found the sink}) {$|S|=1$ {\bf and} $p=0$} {
			\Return {\sc Pop}$(S)$\;
		}
	}
	\Return {\bf nil}\;
}
\caption{Bubble detection}\label{alg:popbub}
\end{algorithm}

After constructing the assembly graph, miniasm removes transitive
edges~\citep{Myers:2005bh}, trims tipping unitigs composed of few reads (4 by default) and pops small
bubbles~\citep{Zerbino:2008uq}. Algorithm~\ref{alg:popbub} detects bubbles where the longest path is shorter than $d$ (50kb by default). It is
adapted from Kahn's topological sorting algorithm~\citep{Kahn62aa}. It starts
from the potential source and visits a vertex when all its incoming edges are
visited before. Algorithm~6 only detects bubbles. We can keep track of the
optimal parent vertex at line~1 and then backtrack to collapse bubbles to a
single path. Fermi~\citep{Li:2012fk} uses a similar algorithm except that it
keeps two optimal paths through the bubble.  \citet{DBLP:conf/wabi/OnoderaSS13}
and \citet{TCS15} have also independently found similar algorithms.

In addition, if $v\to w_1$ and $v\to w_2$ exist and $\ell(v\to w_1)<\ell(v\to
w_2)$, miniasm removes $v\to w_2$ if $[|v|-\ell(v\to w_2)]/[|v|-\ell(v\to
w_1)]$ is small enough (70\% by default). When there are longer overlaps,
shorter overlaps after transitive reduction may be due to repeats.
However, non-repetitive overlaps may also be removed at a small chance, which
leads to missing overlaps and misassemblies.

\subsubsection{Generating unitig sequences}

If there are no multi-edges in the assembly graph, we can use $v_1\to
v_2\to\cdots\to v_k$ to represent a path consisting of $k$ vertices. The
sequence spelled from this path is the concatenation of vertex substrings:
$v_1[1,\ell(v_1\to v_2)]\circ v_2[1,\ell(v_2\to v_3)]\circ\cdots\circ
v_{k-1}[1,\ell(v_{k-1},v_k)]\circ v_k$, where $v[i,j]$ is the substring between
$i$ and $j$ inclusive, and $\circ$ is the string concatenation operator.

In a transitively reduced graph, a \emph{unitig}~\citep{Myers:2000kl} is a path $v_1\to
v_2\to\cdots\to v_k$ such that ${\rm deg}^+(v_i)={\rm deg}^-(v_{i+1})=1$ and i)
$v_1=v_k$ or ii) ${\rm deg}^-(v_1)\not=1$ and ${\rm deg}^+(v_k)\not=1$.
Its sequence is the sequence spelled from the path. Intuitively, a unitig is a
maximal path on which adjacent vertices can be ``unambiguously merged'' without
affecting the connectivity of the original assembly graph.

As miniasm does not correct sequencing errors, the error rate of unitig
sequence is the same as the error rate of the raw input reads. It is in theory
possible to derive a better unitig sequence by taking the advantage of read
overlaps. We have not implemented such a consensus tool yet.

\subsection{Formats: PAF and GFA}

\subsubsection{Pairing mapping format (PAF)}

\begin{table}[tb]
\processtable{Pairwise mapping format (PAF)}
{\footnotesize\label{tab:paf}
\begin{tabular}{rcl}
\toprule
Col & Type & Description \\
\midrule
1 & string & Query sequence name \\
2 & int    & Query sequence length \\
3 & int    & Query start coordinate (BED-like) \\
4 & int    & Query end coordinate (BED-like) \\
5 & char   & `+' if query and target on the same strand; `-' if opposite \\
6 & string & Target sequence name \\
7 & int    & Target sequence length \\
8 & int    & Target start coordinate on the original strand \\
9 & int    & Target end coordinate on the original strand \\
10& int    & Number of matching bases in the mapping \\
11& int    & Number bases, including gaps, in the mapping \\
12& int    & Mapping quality (0--255 with 255 for missing) \\
\botrule
\end{tabular}
}{PAF is TAB-delimited text format with each line consisting of the above fixed
fields. When the alignment is available, column 11 equals the total number of
sequence matches, mismatches and gaps in the alignment. Column 10 divided by
column 11 gives the alignment identity. If the detailed alignment is not
available, column 10 and 11 can be approximate. PAF may optionally have
additional fields in the SAM-like typed key-value format~\citep{Li:2009ys}.}
\end{table}

PAF is a lightweight format keeping the key mapping information (Table~\ref{tab:paf}).
Minimap outputs mappings in PAF, which are taken by miniasm as input for
assembly. We also provide scripts to convert DALIGNER, MHAP and SAM formats to
PAF.

\subsubsection{Graphical fragment assembly format (GFA)}

\begin{table}[tb]
\processtable{Graphical fragment assembly format (GFA)}
{\footnotesize\label{tab:gfa}
\begin{tabular}{clp{5.8cm}}
\toprule
Line & Comment & Fixed fields \\
\midrule
H    & Header  & N/A \\
S    & Segment & segName,segSeq \\
L    & Overlap & segName1,segOri1,segName2,segOri2,CIGAR \\
\botrule
\end{tabular}
}{GFA is a line-based TAB-delimited format. Each line starts with a single
letter determining the interpretation of the following TAB-delimited fields. In
GFA, segment refers to a read or a unitig. A line start with `S' gives the name
and sequence of a segment. When the sequence is not available, it can be a star
`*'. Overlaps between segments are represented in lines starting with `L',
giving the names and orientations of the two segments in an overlap. The last
field `CIGAR' on an `L'-line describes the detailed alignment of the overlap if
available. In addition to the types of lines in the table, GFA may contain
other line types starting with different letters. Each line may optionally have
additional SAM-like typed key-value pairs.}
\end{table}

GFA is a concise assembly format (Table~\ref{tab:gfa}; http://bit.ly/gfaspec) initially proposed by
us prior to miniasm and later improved by community (P. Melsted, S.  Jackman,
J. Simpson and E. Garrison, personal communication). GFA has an explicit
relationship to an assembly graph -- an `S' line in the GFA corresponds to a
vertex and its complement in the graph; an `L' line corresponds to an edge and
its complement. GFA is able to represent graphs produced at all the stages of
an assembly pipeline, from initial read overlaps to the unitig relationship in
the final assembly.

FASTG (http://bit.ly/fastgfmt) is another assembly format prior to GFA.
It uses different terminologies. A vertex in an assembly graph is called an
edge in FASTG, and an edge is called an adjacency. In FASTG, subgraphs can be
nested, though no tools work with nested graphs due to technical complications.
In addition, with nesting, one assembly graph can be represented in distinct
ways, which we regard as a limitation of FASTG.

\subsection{Evaluating the layout accuracy}\label{sec:eval}

Miniasm outputs the approximate positions of trimmed reads on the resulting
unitigs. We extract these reads, map to the true assembly with minimap (option:
`-L100 -m0 -w5') and select the best mapping for each read. For a read $i$, let
${\rm utg}_i$ be the unitig name and ${\rm rank}_i$ be its index on ${\rm
utg}_i$ (i.e. read $i$ is the ${\rm rank}_i$-th read on the unitig).  If two
reads $i$ and $j$ are mapped adjacently on the true assembly, we say the
adjacency is \emph{$w$-consistent}, if (i) ${\rm utg}_i={\rm utg}_j$ and $|{\rm
rank}_i-{\rm rank}_j|<w$, or (ii) both read $i$ and $j$ are the first or the
last $w$ reads of some unitigs. We use $w=5$ to detect large structural
misassemblies.

\end{methods}

\begin{table}[tb]
\processtable{Evaluation data sets}
{\footnotesize\label{tab:data}
\begin{tabular}{llrrr}
\toprule
Name & Species & Size & Cov. & N50 \\
\midrule
PB-ce-40X     & {\it Caenorhabditis elegans}      & 104M & 45  & 16572 \\
ERS473430     & {\it Citrobacter koseri}          & 4.9M & 106 & 7543  \\
ERS544009     & {\it Yersinia pseudotuberculosis} & 4.7M & 147 & 9002  \\
ERS554120     & {\it Pseudomonas aeruginosa}      & 6.4M & 90  & 7106  \\
ERS605484     & {\it Vibrio vulnificus}           & 5.0M & 155 & 5091  \\
ERS617393     & {\it Acinetobacter baumannii}     & 4.0M & 237 & 7911  \\
ERS646601     & {\it Haemophilus influenzae}      & 1.9M & 258 & 4081  \\
ERS659581     & {\it Klebsiella sp.}              & 5.1M & 129 & 8031  \\
ERS670327     & {\it Shimwellia blattae}          & 4.2M & 155 & 6765  \\
ERS685285     & {\it Streptococcus sanguinis}     & 2.4M & 224 & 5791  \\
ERS743109     & {\it Salmonella enterica}         & 4.8M & 188 & 6051  \\
PB-ecoli      & {\it Escherichia coli}            & 4.6M & 160 & 13976 \\
PBcR-PB-ec    & {\it Escherichia coli}            & 4.6M & 30  & 11757 \\
PBcR-ONT-ec   & {\it Escherichia coli}            & 4.6M & 29  & 9356  \\
MAP-006-1     & {\it Escherichia coli}            & 4.6M & 54  & 10892 \\
MAP-006-2     & {\it Escherichia coli}            & 4.6M & 30  & 10794 \\
MAP-006-pcr-1 & {\it Escherichia coli}            & 4.6M & 30  & 8080  \\
MAP-006-pcr-2 & {\it Escherichia coli}            & 4.6M & 60  & 8064  \\
\botrule
\end{tabular}
}{Evaluation data set name, species, reference genome size, theoretical
sequencing coverage and the N50 read length. Names starting with ``MAP'' are
unpublished recent ONT data provided by the Loman lab (http://bit.ly/loman006).
Names starting with ``ERS'' are accession numbers of unpublished PacBio data
from the NCTC project (http://bit.ly/nctc3k). PB-ecoli and PB-ce-40X are PacBio
public data sets sequenced with the P6/C4 chemistry (http://bit.ly/pbpubdat;
retrieved on 11/03/2015). PBcR-PB-ec is the PacBio sample data (P5/C3
chemistry) used in the tutorial of the PBcR pipeline; PBcR-ONT-ec is the ONT
example originally used by \citet{Loman:2015xu}. `pls2fasta --trimByRegion' was
applied to ERS* and PB-ecoli data sets as they do not provide read sequences in
the FASTQ format.}
\end{table}

\section{Results}

\subsection{The accuracy of minimap}

We mapped a human PacBio run ``m130928\_232712\_42213\_*.1.*''
(http://bit.ly/chm1p5c3) with minimap and BWA-MEM~\citep{Li:2013aa}
against GRCh37 plus decoy sequences (http://bit.ly/GRCh37d5).
We started from 23,235 reads (131Mbp), filtered out 7,593 reads (10Mbp) without
$\ge$2kbp BWA-MEM alignments, and further dropped 815 reads (11Mbp) with two or more
$\ge$2kbp chimeric alignments and 598 reads (4Mbp) with mapping quality below 10.
Of the remaining reads, we found only 2.0\% not overlapping the best minimap
mapping of the same read. The majority of them hit to the decoy sequence where
defining the true alignment is challenging as decoy is enriched with incomplete
segments of centromeric repeats. If we exclude hits to the decoy, the
percentage drops to 0.7\%. On this input, minimap is 50 times faster than
BWA-MEM, while finding similar best mapping positions.  This experiment
evaluates both the sensitivity and the specificity of minimap: if minimap had
low sensitivity, it would miss the BWA-MEM mapping completely; if minimap had
low specificity, its best mapping would often be a wrong mapping.

To test the sensitivity for read overlapping, we aligned all reads from
PBcR-PB-ec (Table~\ref{tab:data}) against the reference genome with BWA-MEM,
extracted reads with mapping quality $\ge$10, and identified $\ge$2kb overlaps
between the extracted reads based on their positions on the reference genome.
Minimap finds 93\% of these overlaps. It is more sensitive than MHAP in its
sensitive mode (78\%) but less than DALIGNER (98\%).

\subsection{Assembling bacterial genomes}

We evaluated the performance of miniasm on 17 bacterial data sets
(Table~\ref{tab:data}) with command line `minimap -Sw5 -L100 -m0 reads.fa reads.fa $|$
miniasm -f reads.fa -'. Miniasm is able to derive a single contig per
chromosome/plasmid for all but four data sets: 3 extra $>$50kb contigs for
ERS554120, and 1 extra contig for ERS605484, PBcR-ONT-ec and MAP-006-pcr-1
each. In the dotter plot between the assembly and the reference genome (similar
to Figure~\ref{fig:ce}), no large-scale misassemblies are observed.  We also
applied the method in Section~\ref{sec:eval}. Except ERS473430, the miniasm layouts are
5-consistent with the reference assemblies. For ERS473430, the NCTC project page
claimed the sample has a plasmid. Miniasm gives two contigs, but the NCTC
assembly has one contig only. The difference in layout may be an error in the
NCTC assembly.


We have also run the PBcR pipeline~\citep{Berlin:2015xy}. PBcR requires a spec
file. We took `pacbio.spec' from the PBcR-PB-ec example and `oxford.spec' from
PBcR-ONT-ec, and applied them to all data sets based on their data types. MAP*
data sets only provide FASTA sequences for download. We assigned quality 9 to
all bases as PBcR requires base quality. PBcR assembled all PacBio data sets
without extra contigs longer than 50kb -- better than miniasm. However, on the
ONT data sets, PBcR produced more fragmented assemblies for MAP-006-2,
MAP-006-pcr-1 and MAP-006-pcr-2; the PBcR-ONT-ec assembly is 300kb shorter.

With four CPU cores, it took miniasm 14 seconds to assemble the 30-fold
PBcR-PB-ec data set and 2 minutes to assemble the 160-fold PB-ecoli data set.
PBcR, with four CPU cores, too, is about 700 times slower on PBcR-PB-ecoli and
60 times slower on PB-ecoli.  It is slower on low-coverage data
because PBcR automatically switches to the slower sensitive mode. Here we
should remind readers that without an error correction stage, the contig
sequences generated by miniasm are of much lower accuracy in comparison to
PBcR. Nonetheless, miniasm is still tens of times faster than PBcR excluding
the time spent on error correction.

\subsection{Assembling a C. elegans genome}

\begin{figure}[tb]
\includegraphics[width=.48\textwidth]{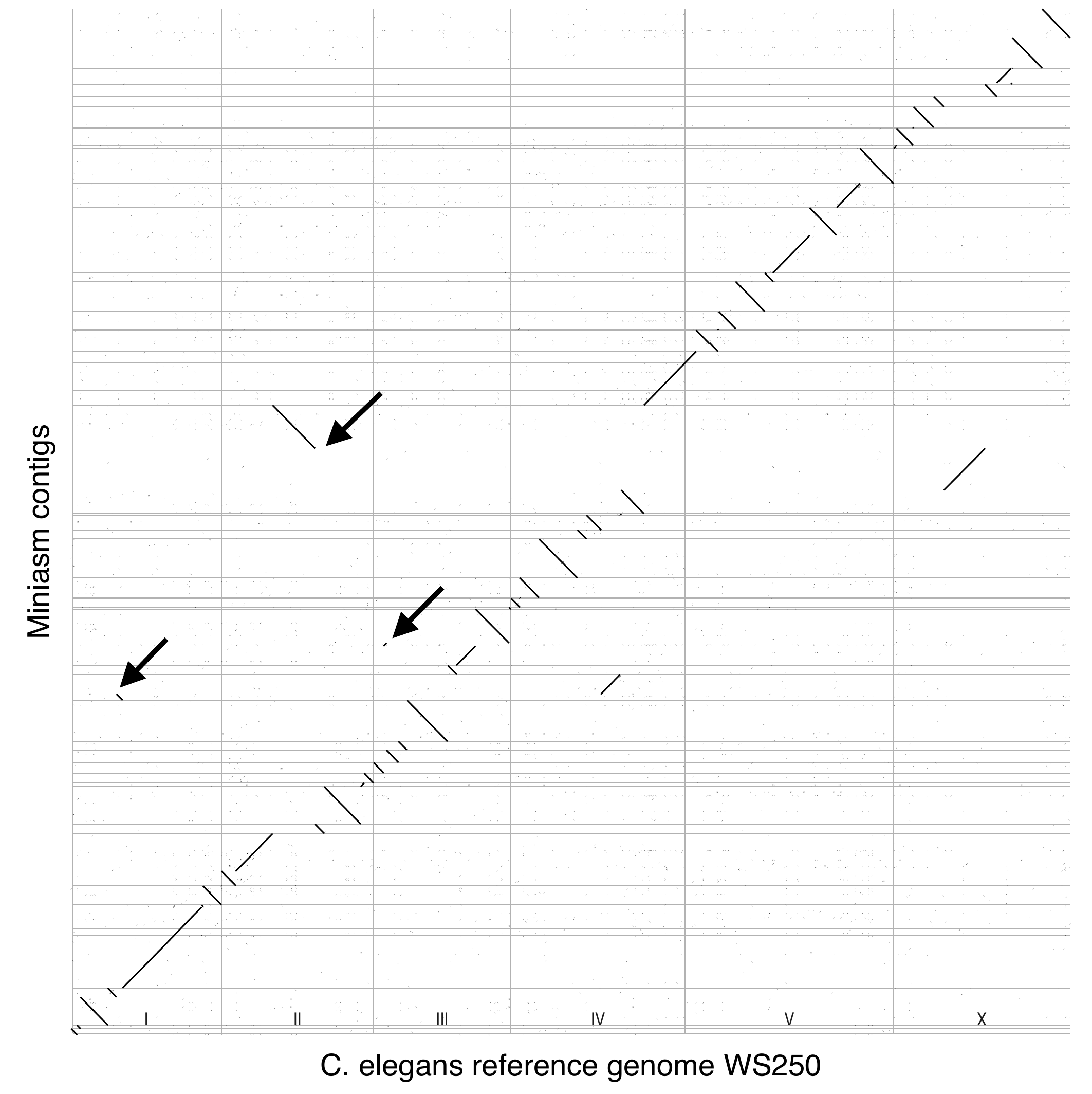}
\caption{Dotter plot comparing the miniasm assembly and the {\it C. elegans}
reference genome. Thin gray lines mark the contig or chromosome boundaries. The
three arrows indicate large-scale misassemblies visible from the
plot. The mapping is done with `minimap -L500'.}\label{fig:ce}
\end{figure}

We assembled a 45-fold {\it C. elegans} data set (Table~\ref{tab:data}). With 16 CPU cores,
miniasm assembled the data in 9 minutes, achieving an N50 size 2.8Mb. From the
dotter plot (Figure~\ref{fig:ce}), we observed three structural misassemblies
(readers are advised to zoom into the vector graph to see the details).
PacBio has assembled the same data set with HGAP3~\citep{Chin:2013qr}. HGAP3
produces shorter contigs (N50=1.6Mb), but does not incur large-scale
misassemblies visible from the dotter plot between the {\it C. elegans}
reference genome and the contigs.

When we take the {\it C. elegans} reference genome as the truth, the method in
Section~\ref{sec:eval} also identifies the three structural misassemblies. The
method additionally finds eight intra-unitig and one inter-unitig
inconsistencies. In all cases, miniasm agrees with HGAP3, suggesting these
inconsistencies may be true structural variations between the reference strain
and the sequenced strain.

We have also tried PBcR on this data set. Based on the intermediate progress
report, we estimated that with 16 CPU cores, it would take a week or so to
finish the assembly in the automatically chosen `sensitive' mode.

For this data set, minimap takes 27GB RAM at the peak. As minimap loads 4Gbp
bases to index, the peak RAM will be capped around 27GB. The memory used by
miniasm is proportional to the number of overlaps.  Although it only takes
1.3GB RAM here, it will become the limiting factor for larger data sets.

\subsection{Switching read overlappers}

Miniasm also works with other overlappers when we convert their output format
to PAF. On the 30-fold PBcR-PB-ec data set, we are able to produce a single
contig with DALIGNER (option -k15 -h50), MHAP (option
\mbox{--pacbio-sensitive}) and GraphMap (option -w owler). DALIGNER is the
fastest, taking 65 seconds with four CPUs.  Minimap is five times as fast on
this data set and is 18 times as fast on PB-ecoli at 160-fold. Minimap is
faster on larger data sets possibly because without staging all possible hits
in RAM, minimap is able to process more reads in a batch while a large batch
usually helps performance. We should note that DALIGNER generates alignments
while minimap does not. Minimap would probably have a similar performance if it
included an alignment step.

\section{Discussions}

Miniasm implements the `O' and `L' steps in the Overlap-Layout-Consensus (OLC)
assembly paradigm. It confirms long noisy reads can be assembled without an
error correction stage, and without this stage, the assembly process can be
greatly accelerated and simplified, while achieving comparable contiguity and
large-scale accuracy to existing pipelines, at least for genomes without
excessive repetitive sequences.  Although without the `C' step, miniasm
cannot produce high-quality consensus for many analyses, it opens the door
to ultrafast assembly if we can develop a fast consensus tool matching the
speed of minimap and miniasm. In addition, MinION has a `read-until' mode,
allowing users to pause sequencing and reload samples. Fast layout by miniasm
could already help to decide if enough data have been collected.

Our main concern with miniasm is that when we look at a low-identity match
between two noisy reads, it is difficult to tell whether the low identity is
caused by the stochastically higher base error rate on reads, or because
reads come from two recent segmental duplications.
In comparison, error correction takes the advantage of multiple reads and in
theory has more power to distinguish high error rate from duplications/repeats.
Bacteria and {\it C. elegans} evaluated in this article are repeat sparse.
We are yet to know the performance of miniasm given repeat-rich genomes.  In addition, miniasm has
not been optimized for large repeat-rich genomes. It reads all hits into RAM,
which may not be practical when there are too many. We need to filter
repetitive hits, introduce disk-based algorithms (e.g. for sorting) or stream
hits before removing contained reads. Working with large complex genomes will
be an important future direction.


Oxford Nanopore is working on PromethION and PacBio will ship PacBio Sequel
later this year. Both sequencers promise significantly reduced sequencing cost and
increased throughput, which may stimulate the adoption of long-read sequencing
and subsequently the development of long-read mappers and assemblers. We hope
in this process, the community could standardize the input and output formats
of various tools, so that a developer could focus on a component he or she
understands best. Such a modular approach has been proved to be fruitful in the
development of short-read tools -- in fact, the best short-read pipelines all
consist of components developed by different groups -- and will be equally
beneficial to the future development of long-read mappers and assemblers.

\section*{Acknowledgement}

We thank P\'all Melsted for maintaining the GFA spec and are grateful to Gene
Myers, Jason Chin, Adam Phillippy, Jared Simpson, Zamin Iqbal, Nick Loman and
Ivan Sovic for their presentations, talks, comments on social media and
unpublished works which have greatly influenced and helped the development of
minimap and miniasm.

\paragraph{Funding\textcolon} NHGRI U54HG003037; NIH GM100233

\bibliography{miniasm}
\end{document}